# A CRYPTOGRAPHIC MUTUAL AUTHENTICATION SCHEME FOR WEB APPLICATIONS

Yassine Sadqi, Ahmed Asimi and Younes Asimi

Information Systems and Vision Laboratory, Department of Mathematics and Computer Science, Faculty of Science, Ibn Zohr University, 8106 City Dakhla Agadir Morocco

*ABSTRACT*

*The majority of current web authentication is built on username/password. Unfortunately, password replacement offers more security, but it is difficult to use and expensive to deploy. In this paper, we propose a new mutual authentication scheme called StrongAuth which preserves most password authentication advantages and simultaneously improves security using cryptographic primitives. Our scheme not only offers webmasters a clear framework which to build secure user authentication, but it also provides almost the same conventional user experience. Security analysis shows that the proposed scheme fulfills the required user authentication security benefits, and can resist various possible attacks.*

*KEYWORDS*

*User authentication, web security, password replacement, cryptography primitives, authentication scheme*

## 1. INTRODUCTION AND NOTATIONS

On the web, user authentication is the mechanism used to validate user's login information. Although authentication lies in the heart of a web application's security, the majority of authentication on the web is built on username/password [1]–[3]. In fact several factors are behind this domination, user experience, simplicity and performance [4].

At first, personal websites employed the built-in basic HTTP authentication specification for password submission. a HTTP server then challenges the user to supply credentials (username and password) in the browser built-in dialog box. If the server validates the login information, the user will be allowed to access the protected resources. The RFC 2617 defines the digest scheme as an extension for HTTP/1.1. It is a security enhancement alternative that avoids the most serious flaws of basic HTTP authentication. However, digest does not provide any security of the content and represents multiple security risks.

Currently with the wide adoption of dynamic web programming, HTTP authentication is rarely used in real-world scenarios [5]. In fact instead of relying on server mechanisms, web applications have become capable of directly validating user's authentication information. For instance, relying on integrated browser dialog boxes do not integrate well with web 2.0 applications that desire to attract the maximum number of users. Therefore, using a HTML form with input fields that allows users to enter their username and password has become the dominant authentication option due to its rich user experience and flexibility [6]. After the user submits the form it is passed through a HTTP method (Get or POST) to the web application. If the credentials are correct, then the user is authenticated.

The lack of a standard form-based authentication and the limited security background of webmasters, has created a set of unique design and implementation choices, which contain various authentication flaws [5], [6], [4]. It is known that passwords often offer poor security and there are numerous publications that have studied these in-depth passwords issues [5], [7], [8]. These problems have led to us to view password-based systems as weak authentication systems, hated by users and have reached their limits [9]–[11]. Security experts recognize that we need a replacement scheme. The National Strategy for Trusted Identities in Cyberspace (NSTIC) declared that "passwords are inconvenient and insecure [12]". In the same direction, the European Central Bank (ECB) issued recommendations for internet payments that require the implementation of strong user authentication [13].

While the security community has proposed a wide range of secure propositions, starting with the TLS client certificate [14] to hardware-based or phone-based multi-factor authentication [10], [15]–[18], Bonneau et al [19] demonstrated in a large study of 35 password replacement options that the majority offer more security than passwords, but they are difficult to use and expensive to deploy. In other words, it is easy to provide security if you do not care about other factors. Furthermore, Grosse and Upadhay [20] mentioned that not all account types need more secure authentication mechanisms. We believe that trying to provide a single solution for authenticating all account types is the ultimate recipient of failure. Therefore, we suggest separating user accounts into three categories:

1) Low-value: Security in this kind of account is not a concern. Webmasters and users are willing to adopt this easy and lightweight authentication.
2) Medium-value: Security is important, but user experience and cost are the determining factors. Because of the market competition, players in this field are willing to use a cost-effective and usable but less secure scheme, compared to a more secure but cost-per-user protocol that alters the user authentication behavior.
3) High-value: Security is the priority, and user experience and performance are not determining factors.

Passwords excel in the first category, but we believe that not all schemes can offer the same set of advantages [21]. For the high-value accounts, webmasters can mandate using hardware-based schemes which score very well in security benefits [19]. Webmasters and users of medium-value category suffer from the absence of a convenient and secure replacement of passwords.

In this paper we propose a new mutual authentication scheme that preserves most username/password authentication advantages and simultaneously improves security using cryptography primitives. Even the most inexperienced user can authenticate without even noticing the background tasks handle by the browser. Compared with the previous propositions, our scheme not only offers webmasters with a clear framework within which to build secure a web user authentication scheme, but it also provides almost the same password user experience which does not require any additional hardware except a modern web browser. Our scheme offers another advantage which is that the user's identity and the associated public key are never stored nor transmitted as plain text. This enhances the security of the proposed protocol and protects the user's anonymity. Our proposed scheme is designed with browser vendors in mind. Native support in browsers will afford improvements in user experience, security and performance. Our security analysis shows that the proposed protocol fulfills the security benefits that a secure user authentication scheme should provide, and can resist various possible attacks. The remainder of this paper is organized as follows. In Section 2, we summarize and discuss related authentication schemes to enhance web user authentication security and present their limits. In Section 3, we give background information on our proposition and identify properties required to provide webmasters and users with a secure and practical authentication. In Section 4, we propose a new user authentication scheme that we call StrongAuth and present its mutual authentication protocol. In Section 5, we discuss the security analysis of the proposed protocol. We conclude the paper in Section 6.

The notations used in this paper are listed in Table 1:

Table 1: Notations

| | |
|---|---|
| $U_i$ | $i^{th}$ User |
| $ID_i$ | Unique identifier in the web application of user $U_i$ |
| $P_i$ | Text secret of user $U_i$ |
| $Salt_i$ | Cryptographic salt generated by the browser for each $P_i$ |
| d | Web application domain name |
| $RW_i$ | Random value used at most once within the scope of a given session generated by the web application for user $U_i$ |
| $RB_i$ | Random value used at most once within the scope of a given session generated by the browser for user $U_i$ |
| $USK_i$ | Private key of user $U_i$ generated by the browser |
| $UPK_i$ | Public key of user $U_i$ generated by the browser |
| SSK | Web Server Private Key |
| SPK | Web Server Public Key |
| encKey | Encryption key generated from $P_i$ using a key derivation function KDF |
| $E_a(b)$ | Encryption of b by a |
| $D_a(b)$ | Decryption of b by a |
| $Sig_{USKi}()$ | Digital signature using the user's $U_i$ private key |
| H( ) | Cryptographic one way hash function. |
| KDF() | Secure key derivation function |
| $SS_i$ | Session key shared between $U_i$ and the web application using HTTPS |
| SK | Fresh symmetric authentication tracking key |
| NIST | National Institute of Standards and Technology |
| ‖ | Concatenation |
| ⊕ | XOR operation |

## 2. RELATED WORK

In this section, we summarize and discuss related authentication methods used in practice or proposed in the literature to enhance password authentication on the web and present their limits.

**Strong password policy**: One of the most deployed cost-efficient techniques to improve passwords security is mandating more difficult to guess passwords. While using this method may provide security against online guessing attacks (dictionary and brute force attacks), it cannot protect users against phishing and key-logging which are two of the major users of authentication attacks [22]. Furthermore, numerous accounts with strong passwords are hard to remember and some argue that from an economic viewpoint, users reject choosing hard to guess passwords [23].

**Two-Factor authentication**: NIST defines three main authentication factors: (1) something the user knows, such as a password or PIN (2) something the user has, such as a smart card or digital certificate (3) something the user is, for example, a fingerprint or other biometric information [24]. Two-factor authentication, or more generally multi-factor authentication, is a form of authentication that relies on at least two-factors. Traditionally, in addition to passwords, most proposed schemes add a smart card as the second factor [25]. Although hardware-based authentication could enhance the security of user authentication on the web, in return there is a big price to pay:

International Journal of Network Security & Its Applications (IJNSA) Vol.6, No.6, November 2014

- *Cost*: even when users care about security, the majority of users may prefer to deal with password risks than buy an additional device.
- *Hardware device management*: users can use password manager software or Single Sign On technology to manage multiple passwords. Nevertheless even with multiple hardware devices are used; they can be easily forgotten or lost.
- *User acceptability*: users are resistant to innovation that alters their behavior [23], thus any complex or additional steps than the conventional username/password are hard to adopt.

As a solution to the above limitations, a wide range of two-factor authentification changed their focus to phone-based as a replacement for hardware dedicated devices [16]–[18]. The following three main assumptions can be made:

- *Cost-efficiency*: almost everybody already owns a cell phone; there is no need to buy an additional device.
- *Usability:* users are familiar with how to use a mobile phone.
- *Availability*: phone is with the user at all times.

While we agree with most of these assumptions, phone-based authentication raises several problems:

- *Security*: mobile usability constraints can make phishing more common in mobile than in Desktop [26]. For instance, it is difficult to know the difference between HTTP and HTTPS URL in a mobile Web browser.
- *Phone-power*: It is known that phone CPU and memory power has widely increased, but in general usage case performance is always less than a personal PC.

**SSL/TLS client authentication** [14]: Both the Secure Socket Layer (SSL) and Transport Layer Security (TLS) provide an optional mechanism to authenticate clients based on public key X509 v3 certificate. Currently this method is the only secure standard for user authentication on the Web. Because of its implementation and administration costs, SSL/TLS client authentication is rarely used on the Web [27], [6]. Additionally, the authentication procedure is complex for non-technical users. Other limitations are discussed in [16], [28].

**PhoneAuth** [16]: PhoneAuth takes a new approach of how to use public key cryptography for strong user authentication on the Web. While this scheme sheds insight on a new design possibility of public key cryptography for user authentication on the web and offers a secure alternative of password, we identified several issues related to the overall solution. As we've discussed, phone-based authentication creates several problems which make it difficult to replace passwords. One of the main issues with this solution is the reliance on connectivity mechanisms that may not be available in certain situations. For example, most personal computers nowadays do not have integrated Bluetooth connectivity. PhoneAuth operations modes present some limitations:

- *Opportunistic mode*: Allows users with a legacy device that can't produce identity assertion or a device with no wireless connection adapter to use traditional password-based login. Although this presents an important usability factor, it will open all the security holes of the traditional password login even if the user does not have a full privilege session.
- *Strict mode*: Though we concede that this mode improves the overall security of authentication on the web, the requirement of the users phone within the proximity of the browser during the first login will create a dependency on a third party device that can be lost at any time.

## 3. BACKGROUND AND GOALS

In this section we will give background information about the ongoing advance of browser-side cryptographic functionalities. Then we will identify properties required to provide webmasters and users with a mutual secure and practical authentication.

### 3.1. Browser Cryptographic Functionalities

Since the web browser is becoming the universal tool for interacting with remote servers, it is natural to ask whether existing browsers can perform cryptographic operations. This need come also with the expanding of Web 2.0 and cloud computing technologies. Thus, we briefly discuss some of in-browser current functionalities.

*Browsers cryptographic libraries*: To support the HTTPS protocol, all modern browsers provide support to some cryptographic operations (e.g. generating the client random certificate and the verify message in the Handshake phase of SSL/TLS protocol [14]). For example, one of the most used cryptographic libraries is Network Security Services [29] which is a set of open source libraries designed to support cross-platform development of security-enabled applications. NSS implements major cryptographic algorithms, and supports smartcards and hardware cryptographic devices using the PKCS#11 module.

*JavaScript cryptography*: In recent discussion of JavaScript cryptography, a controversial issue has been whether or not JavaScript should ever be used for cryptography. On the one hand, the author in [30] strongly argues that it is totally harmful to use JavaScript cryptography inside the browser. However, the authors in [31], [32] argue that claims such as JavaScript crypto isn't a serious research area and is very bad for the advancement of security. While we would agree with the objections against JavaScript cryptography in the context of web browsers as it stands today, we believe that it is possible to improve the language, and core libraries and researchers should work on this in order to enhance the overall security of the web.

*CryptoAPI*: W3C has created the Web Cryptography Working Group to develop a recommendation-track document that defines an API that lets developers implement secure application protocols on the level of Web applications, including message confidentiality and authentication services, by exposing trusted cryptographic primitives from the browser. In fact, as mentioned by the W3C the expected cryptography API recommendation will not be done until January 2015 [33]. The adoption of this standard by web browsers will improves the overall security of web applications. Until this is done, there are two non-standards implementations:

- DOMCrypt [34]: Firefox extension that adds a new JavaScript object to any webpage. DOMCrypt's API is the initial straw man proposal for the W3C's Web Cryptography Working Group.
- PolyCrypt [35]: JavaScript implementation funded by the U.S. Department of Homeland Security that people can use to get a feel for how they can use the CryptoAPI in practice

*Certificate and password managers*: The five most popular browsers (Firefox, Chrome, Internet Explorer, Safari, and Opera) provide certificate management services. Using this built-in functionality, users can display information about the installed certificate including personal and authority certificates that the browser trusts, and perform all the important certificate management actions (import, export, delete). In addition, browser-based password manager is one of the most used techniques to solve passwords memorization problems. The first time the user enters his or her username and password, the browser demands user authorization to store the authentication credentials (username and password) in a special encrypted database. If the user accepts, each time the user visits the same web application, the browser will automatically auto-fill the login form with the appropriate user credentials.

### 3.2. Design Requirements

Learning from previous proposition limitations and the ongoing advance in browser-side functionalities, we identify properties required to provide webmasters and users with a mutual secure and practical web user authentication. We then used these properties to design StrongAuth:

1) *Security:* It will be built on a mechanism that solves password security weaknesses. Our goal is fulfilling the security requirements designed in [19]. In conjunction with SSL/TLS server authentication, the proposed protocol should provide mutual authentication between the user and the web application. User authentication credentials should be stored securely and even with a database compromise, *StrongAuth* should not leak any information.
2) *Usability:* It will provide a similar user experience to the conventional password-based authentication. Even the most inexperienced user can authenticate without even noticing the background tasks handle by the browser.
3) *Adaptability*: Users are resistant for innovation that alters their behavior [23]. Our proposed scheme will include in the registration phase an option that allows activation or deactivation, especially before the wide adoption of our proposed scheme.
4) *Deployability:* Our proposed scheme will require minimal changes in the browser and the web application, and no additional hardware will be required.
5) *Cost-efficiency:* Cost is always a factor that plays a decisive role in real-world scenario. Therefore our scheme will not involve superfluous cost per user, but instead be open-source to implement and deploy by using existing technologies and standards.
6) *Browser support:* Will be implemented as part of the browser (core component or extension) to provide adequate security and functionality guarantees.

## 4. THE PROPOSED SCHEME

In this section, we present our scheme based on the designed properties introduced in Section 3. First we give an overview of our scheme and then present its proposed protocol most important phases: registration phase, login and authentication phase as well as key renewal phase.

### 4.1. Overall Design of StrongAuth

As described in Figure 1, StrongAuth is a decentralized architecture. In fact, each web application manages its own users. This choice is based on the success of this decision in the case of passwords (each application stores passwords of its users), and the failure of the centralized X509 certificates model (the dream of a global public key infrastructure never emerged because of several constraints, N. Ferguson and B. Schneier [36] gave comprehensible explanations about PKI issues).

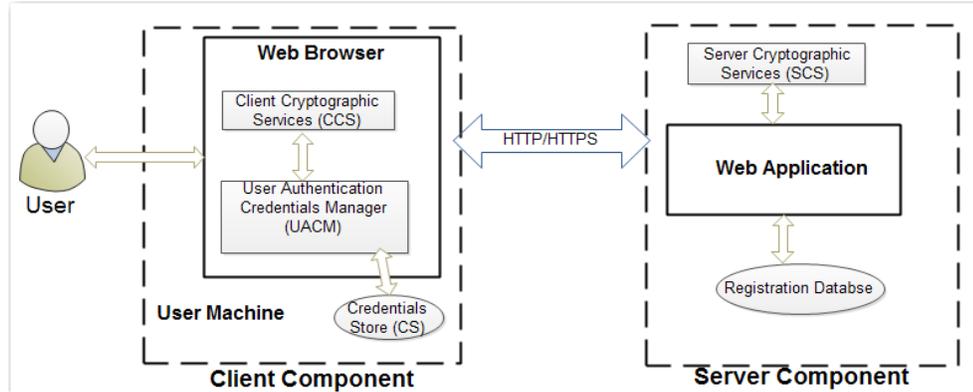

Figure 1. StrongAuth Proposed Design

Our overall scheme is split into client a component side and a server component side. For each user, the client component includes:

a. Web browser that support HTTPS.
b. *Client Cryptographic Services (CSS)* implements the browser-side *StrongAuth* cryptographic operations like the:
    o Regeneration of pseudo-random sequences;
    o Creation of asymmetric key pair;
    o Generation of symmetric key using a key derivation function;
    o Encryption of keys;
    o Creation and validation of digital signature;
c. *User Authentication Credentials Manager (UACM)* is the core of our proposition. Similar to modern browsers certificate and password managers the UACM is responsible for:
    o Storing user's credentials in the credential store.
    o Importing and exporting user's credentials.
d. *Credentials Store (CS)* is typically a persistent storage database. The basic format of a record is shown in Table 2. It includes four fields, but can be further extended if necessary:
    o *Identification* represents the primary key of the table. For each $U_i$ a unique *ID* exists in a specific web application. To protect the user identity $ID_i$ confidentiality, the browser store $ID_i$ hash value concatenated with *d*. In case of CS compromise the attacker will only get the hash value: $H(ID_i || d)$.
    o *Private key* is the user $Ui$ encrypted private key using *encKey* which is generated from $P_i$ and $Salt_i$ using a secure key derivation function and a NIST-approved symmetric algorithm: $E_{encKey}(USK_i)$
    o *Salt* is a large and sufficiently pseudo-random salt generated by the web application. It is used to calculate the encryption key *encKey* for encrypting the $USK_i$.
    o *Public key* contains the encrypted format of the user $U_i$ public key using: $E_{encKey}(UPK_i)$.

Table 2: The basic format of a user authentication credential

| *Identification* | *Private key* | *Salt* | *Public key* | … |
|---|---|---|---|---|

The server component contains the following:

- *Server Cryptographic Services (SCS)* works as the web application engine for the registration, authentication and key renewal cryptographic operations.
- *Registration Database (RD)* stores after a successful registration the user information that will be used in the authentication and key renewal phases. It contains three field and can be further extended:
    - *Identification* stores the user identity $ID_i$ as $B_i = H(ID_i||d)$ to recognize a specific user $U_i$ and simultaneously protect its confidentiality.
    - *Public Key* contains the user $U_i$ public key $UPK_i$, which is stored as $C_i = (UPK_i \oplus SSK \oplus RW_i)$ to preserve its secrecy.
    - *Random* is stored as $SR_i = RW_i \oplus H(SSK)$ and used to retrieve the $UPK_i$

The process of our scheme includes three phases: registration phase, login and authentication phase and the key renewal phase. The steps of our proposed protocol are depicted in Figure 2.

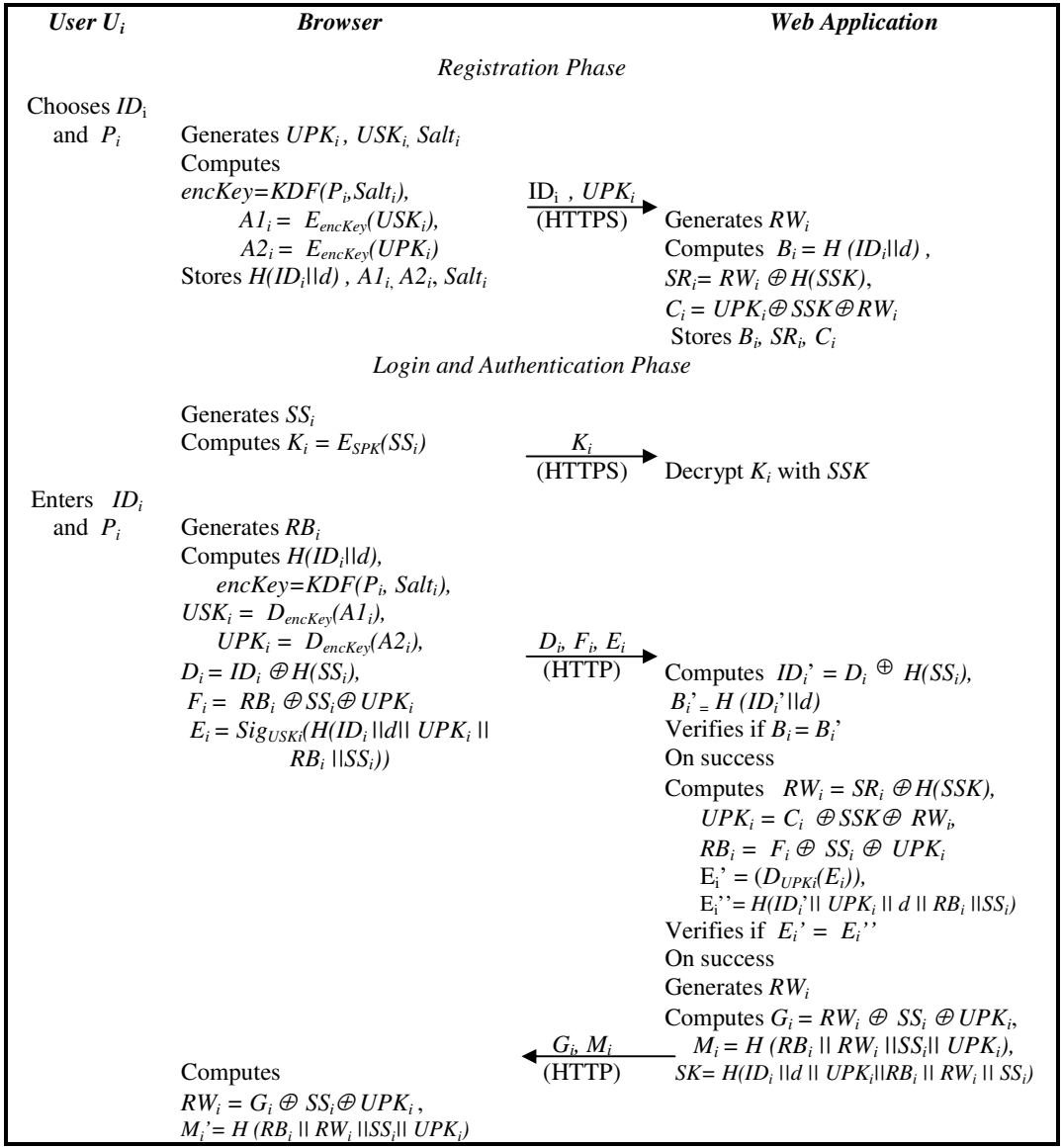

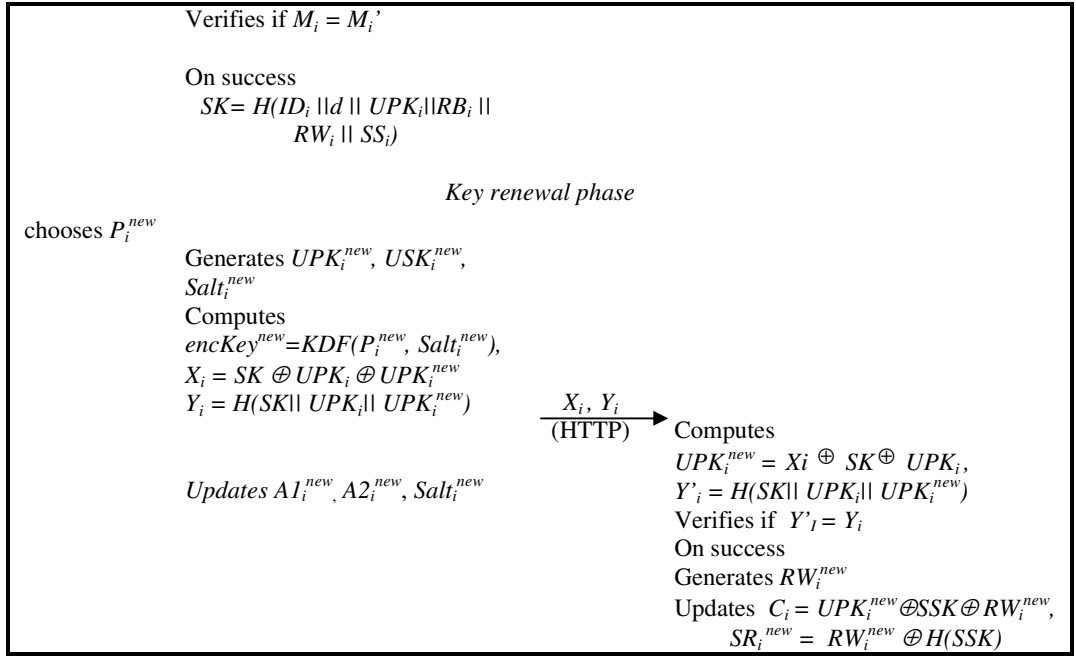

Figure 1. StrongAuth proposed Protocol

### 4.2. Registration Phase

A new user has to register to the web application to become a legal client. The protocol at first uses HTTPS to provide server authentication using public key certificate and to protect the confidentiality and integrity of the user's identity $ID_i$ and the associated public key $UPK_i$.

1) Similar to traditional password authentication mechanism, the user $U_i$ chooses a user identifier $ID_i$ and a text secret $P_i$. If the user chooses to activate *StrongAuth*, the browser executes the following background actions:
    a. Generates an asymmetric key pair $UPK_i$, $USK_i$ and $Salt_i$
    b. Sends the user $ID_i$ and the associated public key $UPK_i$ over a secure HTTPS session to the web application.
    c. Computes the symmetric encryption key $encKey$ from the user $U_i$ secret $P_i$ and $Salt_i$ using a secure key derivation functions $encKey=KDF(P_i, Salt_i)$. To improve the security and usability of $P_i$, we could give $U_i$ the possibility to generate a strong text secret from a digital personal object [37].
    d. Computes $A1_i = E_{encKey}(USK_i)$, $A2_i = E_{encKey}(UPK_i)$, and stores in its database the user authentication credentials : $H(ID_i||d)$, $A1_i$, $A2_i$, $Salt_i$.
2. The web application generates random value $RW_i$. Next, it calculates and stores in its database $B_i$ as $B_i = H(ID_i||d)$, $C_i$ as $C_i = UPK_i \oplus SSK \oplus RW_i$ and $SR_i$ as $SR_i = RW_i \oplus H(SSK)$.

### 4.3. Authentication and Login Phase

When the user $U_i$ wants to login to a web application that implements StrongAuth, his/her browser establishes a secure session with the web server using the HTTPS protocol and requests the authentication service. HTTPS is required to provide server authentication using public key certificate (strong server-side authentication) and to protect the confidentiality and integrity of the session key $SS_i$.

1) After the validation of the server authentication certificate, the browser generates a new SSL/TLS session key $SS_i$, encrypts its using the server public key $SPK$ as $K_i = E_{SPK}(SS_i)$ and sends it to the web application.
2) The server decrypts $K_i$ with its private key $SSK$. After that, all the subsequent messages of this phase are exchanged without using the HTTPS protocol.
3) The user $U_i$ enters his/her identity $ID_i$ and $P_i$ to the browser.
4) The browser computes $H(ID_i||d)$ and retrieves from the $CS$ the matching user credentials $(A1_i, A2_i, Salt_i)$ in case of a correct $ID_i$. If not, the operation is aborted with an error message.
5) The browser generates $RB_i$ and computes $encKey=KDF(P_i, Salt_i)$, $USK_i = D_{encKey}(A1_i)$, $UPK_i = D_{encKey}(A2_i)$, $D_i = ID_i \oplus H(SS_i)$, $F_i = RB_i \oplus SS_i \oplus UPK_i$ and a digital signature using the user $USKi$ as $E_i = Sig_{USKi}(H(ID_i||d||UPK_i||RB_i||SS_i))$. Then, the browser sends $D_i, F_i, E_i$ to the web application.
6) To identify the user $U_i$, the web application computes $ID_i'$ from $D_i$ as $ID_i' = D_i \oplus H(SS_i)$ and $B_i' = H(ID_i'||d)$. If $B_i'$ exists in the registration database, the web application recognize the user $U_i$ and retrieves form its database the corresponding $RW_i$ as $RW_i = SR_i \oplus H(SSK)$ and $UPK_i$ as $UPK_i = C_i \oplus SSK \oplus RW_i$. If not, the request is rejected. Afterwards, the web applications calculates $RB_i$ as $RB_i = F_i \oplus SS_i \oplus UPK_i$, $E_i' = (D_{UPKi}(E_i))$, $E_i'' = H(ID_i'||UPK_i||d||RB_i||SS_i)$. If both $E_i''$ and $E_i'$ values are equal, the user $U_i$ is successfully authenticated and proceeds to the next steps. Otherwise, the login request from the user $U_i$ is rejected.
7) The web application generates $RW_i$ and calculates $G_i = RW_i \oplus SS_i \oplus UPK_i$ and $M_i = H(RB_i||RW_i||SS_i||UPK_i)$, and sends $G_i$ and $M_i$ to $U_i$ browser.
8) The browser calculates $RW_i = G_i \oplus SS_i \oplus UPK_i$, because it knows the public key $UPK_i$ of user $U_i$. After that, the browser computes $M_i'$ as $M_i' = H(RB_i||RW_i||SS_i||UPK_i)$, and verifies if $M_i$ value equal $M_i'$. If both values are equal the $U_i$ browser is assured that the messages are sent by the legal web application. Therefore, the mutual authentication between the browser and the web application is successfully accomplished.
9) After successful mutual authentication, both the browser and the web application computes a fresh symmetric key $SK = H(ID_i||d||UPK_i||RB_i||RW_i||SS_i)$, which will then serve as the basis for authentication tracking.

### 4.4. Key Renewal Phase

In password-based authentication, it is recommended to change the password regularly. Similarly, cryptographic keys should have a definite lifetime. In our protocol, the process from the user $U_i$ perspective is similar to changing his/her password. $U_i$ browser's takes charge of the additional steps.

1. The user $U_i$ authenticates to the web application.
2. After successful mutual authentication between the user $U_i$ and the web application, $U_i$ chooses a new $P_i^{new}$.
3. The browser generates new user credentials $UPK_i^{new}$, $USK_i^{new}$, $Salt_i^{new}$, $A1_i^{new}$, $Salt_i^{new}$ and sends to the web application $X_i = SK \oplus UPK_i \oplus UPK_i^{new}$ and $Y_i = H(SK||UPK_i||UPK_i^{new})$. After that, the browser computes and updates its database: $A1_i = E_{encKey}(USK_i^{new})$, $A2_i = E_{encKey}(UPK_i^{new})$, $Salt_i^{new}$.
4. The web applications computes $UPK_i^{new}$ from $X_i$ as $UPK_i^{new} = Xi \oplus SK \oplus UPK_i$ and verifies if $Y'_i = H(SK||UPK_i||UPK_i^{new})$ value equals the received $Y_i$ value. If both values are equal, the web application is assured that the request is sent by the legitimate user $U_i$ and the message is not tampered during transmission. As a result, the web application generate a $RW_i^{new}$ and updates $C_i = UPK_i^{new} \oplus SSK \oplus RW_i^{new}$, $SR_i^{new} = RW_i^{new} \oplus H(SSK)$. Otherwise, the request is rejected.

### 4.5. Particular Issues

Without the stored user credentials, a specific user $U_i$ cannot authenticate to his/her web account. In our actual proposition scheme, the authentication information used in the registration phase is stored locally in the user's machine which creates a portability issue. Furthermore, users could lose the authentication credentials by formatting their machine or replacing them with newer modes. Thus, a recovery mechanism is needed. These issues are complicated and have non-trivial tensions between availability, security, usability, accessibility, and privacy. We are working on the possibility of using cloud computing technology, but we are in the beginning stages of building a system that attempts to navigate these various constraints.

Inspiring from a modern browser certificate manager, we propose an importation/exportation mechanism that allows users to backup their authentication information on personal devices that users consistently carry with them (e.g. USB flash memory).

## 5. SECURITY ANALYSIS

In this section, security analysis of the proposed protocol is given. We will prove that the proposed protocol fulfills the eleven security benefits that a secure web user authentication scheme on the web should provide [19], and show that the proposed protocol can resist other possible attacks.

- B1: Resilient-to-Physical-Observation.
- B2: Resilient-to-Targeted-Impersonation.
- B3: Resilient-to-Throttled-Guessing.
- B4: Resilient-to-Unthrottled-Guessing.
- B5: Resilient-to-Internal-Observation.
- B6: Resilient-to-Leaks-from-Other-Verifiers.
- B7: Resilient-to-Phishing.
- B8: Resilient-to-Theft.
- B9: No-Trusted-Third-Party
- B10: Requiring-Explicit-Consent.
- B11: Un-linkable.

### 5.1. Security Benefits

Our scheme is very different compared to others. The user identity $ID_i$ and the associated public key $UPK_i$ both can be considered as a secret parameter of the user because they are never transmitted nor stored as plain text in our scheme. The security of our scheme is not depended on this assumption.

In the proposed protocol, to authenticate the attacker has to generate $D_i = ID_i \oplus H(SS_i)$, $F_i = RB_i \oplus SS_i \oplus UPK_i$, $E_i = Sig_{USKi}(H(ID_i ||d||UPK_i|| RB_i ||SS_i))$, which are completely different with each new user authentication. For instance, the digital signature varies constantly and a new $SS_i$ and $RB_i$ values are regenerated for each new HTTPS session. In addition both $P_i$ and $USK_i$ are never transmitted over the network. Thus, our scheme meets *B1, B2, B3,* and *B4*.

Our proposed protocol fulfills *B5* and *B8*, because we assume that the attacker cannot steals both the $U_i$ private key and the associated $P_i$. In addition, the user $ID_i$ and $UPK_i$ are never stored nor transmitted in a plain text, and the private key $USK_i$ and $UPK_i$ are stored in an encrypted format using a strong NIST approved symmetric algorithm. For more critical application, the private key can be moved out of reach of thieving malware on the user $U_i$ devices, using a hardware Trusted Platform Module (TPM) [38]. It also meets *B6* because even if the attacker steal the verification table by breaking into the web application's registration database, the attacker does not have sufficient information to calculate $B_i = H(ID_i||d)$, $SR_i = RW_i \oplus H(SSK)$, $C_i = UPK_i \oplus SSK \oplus RW_i$, the user's $USK_i$, $ID_i$ and $P_i$. In addition, our proposition is secure against phishing attack (*B7*) since the falsified web application cannot produce valid credentials ($G_i = RW_i \oplus SS_i \oplus UPK_i$, $M_i = H(RB_i || RW_i ||SS_i|| UPK_i)$).

Our proposed protocol requires a trusted-third-party for the server SSL/TLS certificate authentication, which mean that in case of a server certificate authority compromise the attacker can get the session key value $SS_i$ and $ID_i$ from $D_i$ as $ID_i = D_i \oplus H(SS_i)$, but cannot compute a valid $F_i = RB_i \oplus SS_i \oplus UPK_i$, $E_i = Sig_{USKi}(H(ID_i ||d|| UPK_i || RB_i ||SS_i))$ because it does not know the user's $U_i$ private/public key ($USK_i$, $USK_i$), the user $ID_i$ nor the associated user $P_i$.

Our proposed protocol cannot be started without the user $U_i$ submitting to his/her browser the associated secret text $P_i$ to decrypt the corresponding private key $USK_i$, and $UPK_i$ which offers *B10*. The decentralize characteristic of our scheme, and the dynamic identity used in each session $D_i = ID_i \oplus H(SS_i)$ protects the user privacy and makes it unlinkable (*B11*).

## 5.2. Other Attacks

**Eavesdropper:** In this type of attack the adversary can see all the network traffic between the user $U_i$ and the web application, but cannot modify any message. Thus the attacker can reply authenticators. The user $U_i$ never sends his/her private key and text secret $P_i$ over the network. In addition the browser uses pseudo-random nonce $RB_i$ value and $SS_i$ session key to generate dynamic identity and a different signature for each new HTTPS session. Hence our protocol can withstand this attack.

**Message modification or insertion attack:** Message modification and insertion attack gives an attacker the possibility to alter and add some specific information with the attention to gain unauthorized access to the user $U_i$ account. The attacker can alter $SS_i$, $D_i = ID_i \oplus H(SS_i)$, $F_i = RB_i \oplus SS_i \oplus UPK_i$, and $G_i = RW_i \oplus SS_i \oplus UPK_i$, however with the usage of secure digital public key signature and collision free cryptographic hash function our solution is secure against these type of attacks.

**SSL/TLS Man-in-the-middle attack:** It is known that the security of HTTPS depends on correct certificate server authentication signed by a trust certification authority. Unfortunately with current browsers PKI trust-model, hundreds of certification authorities are installed and trusted from more than 50 different countries which results in several reported attacks against this authorities [39]. In this specific attack the attacker has a forged SSL/TLS certificate for the server to which the user $U_i$ is authenticating, thus allowing him to perform SSL/TLS man-in-the-middle attacks. The attacker in this case can fake the user $U_i$ to send the chosen session key $SS_i$ encrypted with the attacker public key. We assume a secure HTTPS server authentication in the registration phase, hence the attacker cannot impersonate the user's $Ui$ since he does not possess the $U_i$ associated private key $USK_i$ or the symmetric encryption key $encKey$. Even with the knowledge of $SS_i$, the adversary cannot computes $E_i' = D_{UPKi}(E_i)$, $E_i'' = H(ID_i ||d ||UPK_i ||RB_i || SS_i)$, $G_i = RW_i \oplus SS_i \oplus UPK_i$, $M_i = H(RB_i || RW_i ||SS_i|| UPK_i)$. Therefore our proposed protocol can withstand SSL/TLS Man-in-the-middle attack.

**Off-line dictionary and brute force attack:** To launch off-line brute force attack, an attacker first steals the user credentials authentication from the user credentials store: $A_i = E_{encKey}(USK_i)$, $A2_i = E_{encKey}(UPK_i)$ and the hash value of the user identity $ID_i$. Even after having an $A_i$, $A2_i$ the attacker has to decrypt both the private and public keys which is not possible without the knowledge of the symmetric encryption key *encKey* since we assume the usage of secure key derivation function with at least 64 bits salt $Salt_i$ long and a minimum of 1000 iteration counts [40]. Also the attacker needs to know the user $ID_i$, which is stored in a hash value using a cryptographic one way hash function. Therefore our protocol is secure against off-line brute force attacks. For additional security we could leverage a Trusted Platform Module (TPM). The user authentication information would reside in the TPM's trusted storage facility, but this is not required in our scheme.

## 6. CONCLUSION

In this paper, we proposed a new mutual authentication scheme that preserves most username/password actual authentication advantages and simultaneously improves security using cryptographic primitives. Indeed, even the most inexperienced user can authenticate without even noticing the background tasks handle by the browser. Compared with the previous propositions, our scheme not only offers webmasters a clear framework within which to build secure web user authentication scheme, but also provides almost the same traditional user experience which does not require any additional hardware. Our scheme has another advantage that the user's identity and the associated public key are never stored nor transmitted as a plain text which enhances the security of the proposed protocol and protects the user's anonymity. Our proposed protocol is designed with adoption by browser vendors in mind. Native support in browsers will afford improvements in user experience, security and performance. Security analysis showed that the proposed protocol fulfils the security benefits that a secure user authentication scheme should provide and can resist to various possible attacks such as SSL/TLS man-in-the-middle attack, off-line dictionary and brute force attack, and message modification or insertion attack. Future scope in this work is to investigate the possibility of using cloud computing technology to improve the portability and recovery of our scheme. We also plan to provide a secure design for user's session management. While the proposed scheme needs public key cryptography, then we will further focus on elliptic curve cryptography which offers faster computations and less power use.

**Authors**

**SADQI Yassine** received his Master's degree in the field of Computer Science and Distributed Systems at Ibn Zoher University in 2012. He is currently a Ph.D. candidate of the Ibn Zoher University, Agadir, Morocco. His main field of research interest is Web Applications Security, Computer Security and Cryptography.
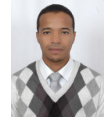

**ASIMI Ahmed** received his PhD degree in Number theory from the University Mohammed V – Agdal in 2001. He is reviewer at the International Journal of Network Security (IJNS). His research interest includes Number theory, Code theory, and Computer Cryptology and Security. He is a full professor at the Faculty of Science at Agadir since 2008.
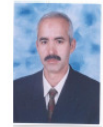

**ASIMI Younes** Received his Master's degree in Computer Science and Distributed Systems in 2012 from Departments of Mathematics and Computer Science, Faculty of Science, University Ibn Zohr, Agadir, Morocco. He is currently pursuing Ph.D in Departments of Mathematics and Computer Sciences, Information Systems and Vision Laboratory, Morocco. His research interests include Authentication Protocols, Computer and Network Security and Cryptography.
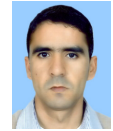